# THE EUCLID VIS READ-OUT SHUTTER UNIT: A LOW DISTURBANCE MECHANISM AT CRYOGENIC TEMPERATURE

C. Larchevêque [2], L. Genolet [1], E. Bozzo [1], S. Paltani [1], D. Manzoni [2], M. Castelli [2], J-P. Heurteau [2], C. Thomas [2], J. Constant [2], R. Michaud [2], A. Marrei [2], N. Martini [2], T. Verhegge [2]

[1] Department of Astronomy, University of Geneva, Chemin d'Ecogia 16, 1290, Versoix, Switzerland, Email: Enrico.Bozzo@unige.ch

[2] APCO Technologies, Chemin de Champex 10, 1860, Aigle, Switzerland, Email: c.larcheveque@apco-technologies.eu

## ABSTRACT

Euclid is the second medium-size mission (M2) of the ESA Cosmic Vision Program, currently scheduled for a launch in 2020. The two instruments on-board Euclid, VIS (VISible imager) and NISP (Near Infrared Spectrometer and Photometer), will provide key measurements to investigate the nature of dark energy, advancing our knowledge on cosmology. We present in this contribution the development and manufacturing status of the VIS Read-out Shutter Unit, whose main function is to prevent direct light from falling onto the VIS CCDs during the read-out of the scientific exposures and to allow the dark-current/bias calibrations of the instrument.

## MISSION

Euclid [1] is the second medium-size mission (M2) of the ESA Cosmic Vision Program, currently scheduled for a launch in 2020.

The mission will investigate the distance-redshift relationship and the evolution of cosmic structures out to redshifts ~2, mapping the geometry of the dark Universe over the entire period during which the dark energy played a significant role in the cosmic acceleration.

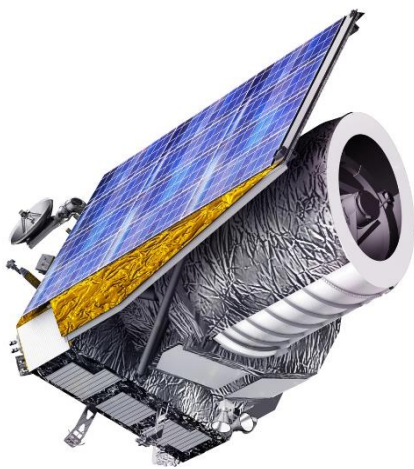

*Figure 1: Artist's impression of the Euclid spacecraft (credits:* ESA).

To accomplish the Euclid mission ESA has selected Thales Alenia Space for the construction of the satellite and its Service Module and Airbus Defence and Space for the Payload Module.

Euclid will be equipped with a 1.2 m diameter Silicon Carbide (SiC) mirror telescope manufactured by Airbus Defence and Space. The science beam from the telescope will illuminate the VIS and NISP instruments, provided by the Euclid Consortium.

The development of the VIS instrument is led by the MSSL (Mullard Space Science Laboratory) and its main goal is to image all galaxies of the Euclid survey with an unprecedentedly high image quality. It will be used to measure the shapes of galaxies and derive the gravitational lensing effects induced on them by large scale structures populating the cosmological Universe. The different subsystems of the VIS instruments are shown on Figure 2.

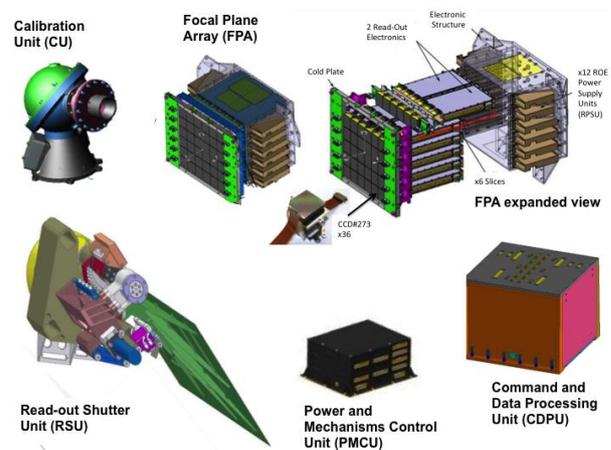

*Figure 2: Overview of the VIS subsystems (courtesy EC / VIS team [1]).*

We present in this contribution the development and manufacturing status of the VIS Read-out Shutter Unit (RSU), whose main function is to prevent direct light from falling onto the VIS CCDs during the read-out of the scientific exposures and to permit the dark-current/bias calibrations of the instrument.

**DESIGN REQUIREMENTS**

Several major challenges have been faced during the development of the RSU due to the harsh conditions of the environment in which the unit will be operated and the strict performance requirements imposed by the needs of the Euclid mission.

First of all, the RSU should retain its performances in vacuum over a large range of temperatures, spanning from room 293 K down to 110 K. The low operating temperatures of the RSU are dictated by the proximity of the unit to the VIS focal plane, the latter being equipped with CCDs (charge-coupled device) that reach the highest performances only in cold conditions [2]. This proximity also imposes strict requirements on the cleanliness of the unit and on the stray-light it can induce on the instrument focal plane.

The second challenging design aspect is related to the high number of actuations to be performed without any failure, exceeding 1 million when the qualification factor is convolved with the expected number of openings and closures of the RSU during the 6.25 years of Euclid scientific observations.

Finally, the RSU is also requested to limit the exported torque and the induced micro-vibrations on the Euclid spacecraft down to an exquisitely low level in order not to degrade the performances of the VIS and NISP instruments. As the motion of the RSU is too short (between 3 and 10 s) to be actively compensated through the usage of the spacecraft attitude and orbit control system (reaction time ~60s), its exported torque and micro-vibrations have to be minimized through a carefully fine-tuned design of the unit.

**RSU DESCRIPTION**

The RSU main function to occult the light beam is achieved through the rotation of its relatively large aluminium leaf by 70 deg (see Figure 3). The rotation is driven by a SAFRAN-SAGEM stepper motor with 360 steps per revolution.

To counteract the moment created by the rotation of the leaf, a spur gear is introduced to drive the movement of a flywheel (FW) in the opposite direction.

The hatch shaft (HS), holding the leaf, and the motor shaft (MS), holding the motor and the FW, are mounted on a titanium alloy base plate through dedicated ball bearings. Two pairs of ball bearings in face-to-face configuration are used for the HS, while a single pair in back-to-back configuration is considered for the MS.

Cold and warm redundant end-switches are installed at the tip of the hatch shaft to indicate when the RSU is in the open or closed position.

Some more details of the design are discussed in the following sections.

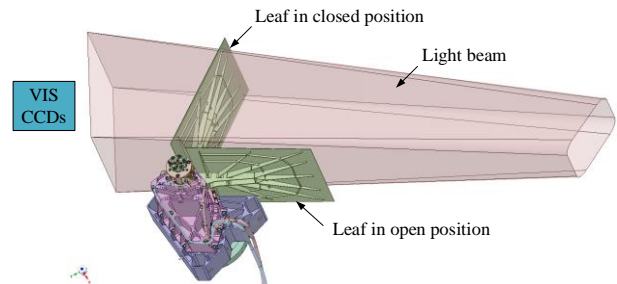

*Figure 3: The RSU in both the open and closed positions intersecting the science beam.*

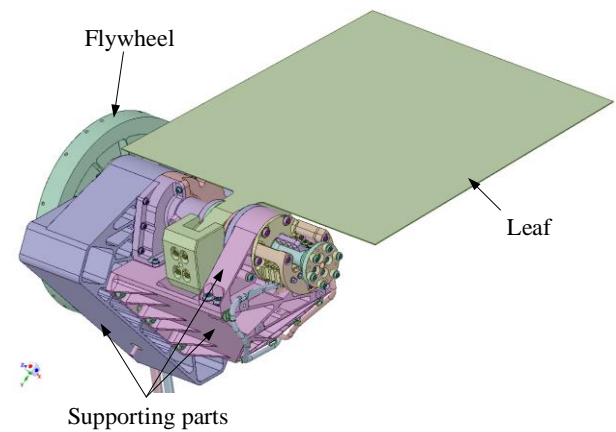

*Figure 4: An overview of the RSU design.*

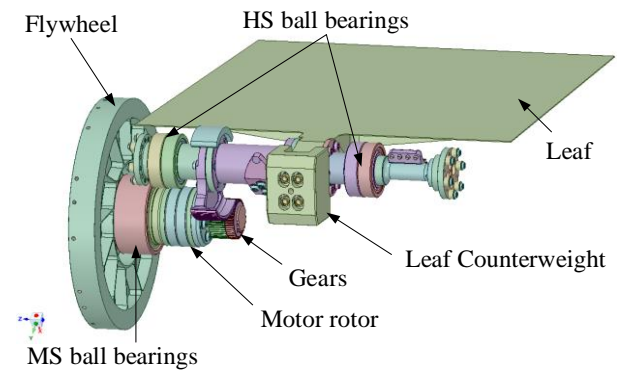

*Figure 5: Details of the RSU mobile parts.*

**KEY FEATURES**

To fulfil all RSU requirements mentioned before, a number of key features have been implemented in the unit design as detailed below.

**Depointing**

To limit the low frequency (<10Hz) component of the

RSU exported loads and minimize the depointing of the Euclid spacecraft, the moment of inertia (MoI) of both the leaf and the flywheel shaft (rotating in opposite directions) are measured through a moment of inertia test stand, developed by Schenck S.A.S. The achievable accuracy in the measurements performed with this machine is as low as 0.1% and adjustment masses have been implemented on the RSU MS side to fine-tune its MoI (see Figure 6).

To optimally compensate for the leaf rotation, the MS and HS should be precisely aligned and thus the final machining of the houses of all ball bearings is carried out once all relevant supporting parts are assembled.

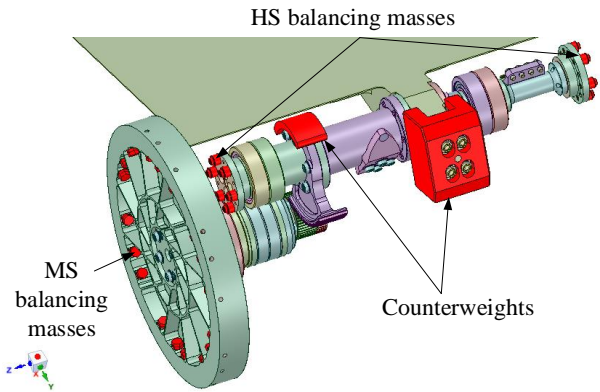

*Figure 6: Components introduced to perform the fine-tuned adjustments of the RSU CoG, MoI, and PoI.*

The counter-balancing of the torque around the rotation axis is not sufficient, as the product of inertia terms in the equation 1 below contribute to generate exported torques also along the XY directions.

$$\begin{pmatrix} I_{xx} & I_{xy} & I_{xz} \\ & I_{yy} & I_{yz} \\ & & I_{zz} \end{pmatrix} \cdot \begin{pmatrix} 0 \\ 0 \\ \ddot{\theta}_z \end{pmatrix} = \begin{pmatrix} I_{xz} \cdot \ddot{\theta}_z \\ I_{yz} \cdot \ddot{\theta}_z \\ I_{xz} \cdot \ddot{\theta}_z \end{pmatrix} \quad (1)$$

To limit the impact of the cross inertia matrix terms the design of the leaf has been carefully symmetrized and both shafts have been equipped with adjustable masses to minimize both the static and dynamic unbalances.

**Thermal stability**

All the adjustments mentioned before are done at room temperature but shall retain their validity across the entire unit operational temperature range (110-160 K). This is ensured by a careful design of all the RSU mobile parts. As an example, the leaf mass is counter-balanced by a large C-shaped tungsten alloy weight (as shown on Figure 6). This shape allows to have the lever between the rotation axis and the centre of gravity (CoG) of the counterweight (CW) in aluminium alloy. This way, the mass ratio equal the levers ratio independently from the temperature and the global CoG of the assembly is maintained on the rotation axis. This is illustrated in Figure 7 and Eq. 2:

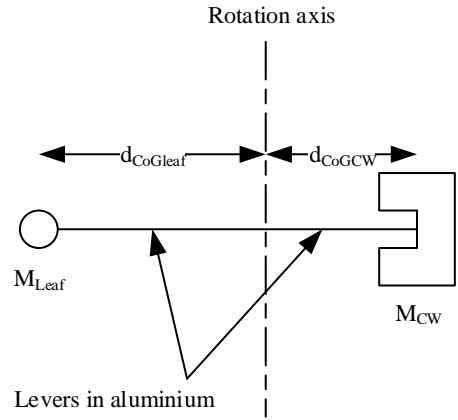

*Figure 7: The principle of the thermal stability.*

$$\frac{M_{leaf}}{M_{CW}} = \frac{d_{CoG_{CW}}}{d_{CoG_{leaf}}} = Cst \quad (2)$$

The changes in the MoI of one shaft with respect to the other due to the thermal shrinking are also compensated by the RSU design, having been first quantified by analysis and then taken into account to optimize the shape of the FW.

The two shafts and all the supporting parts are made of titanium alloy to limit any thermo-elastic distortion.

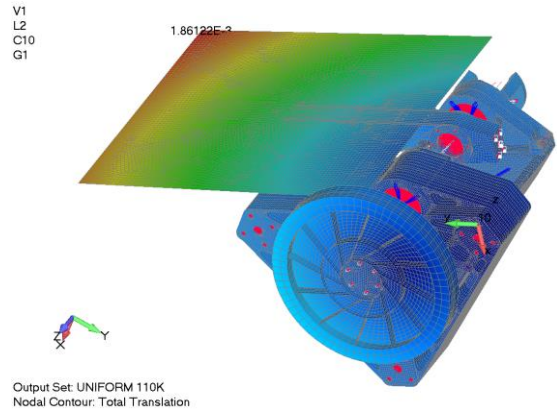

*Figure 8: Thermo-elastic displacements.*

**Micro-vibrations**

The RSU is also expected to limit the "high" frequency (10-500Hz) content of the exported loads to reduce to a negligible level the disturbances to the VIS and NISP instruments. To investigate this aspect, a dedicated model was developed in MATLAB® and SIMULINK® and we identified the following contributions to the RSU high frequency content of the exported loads:

- The closed-loop current and the micro-stepping

- parameters of the motorcontroller;
- The resistance, inductance, detent torque, back electromotive force constant, torque constant, and hysteresis torque factor of the motor;
- The gear ratio, backlash, variable mesh stiffness, friction and damping;
- The friction of the ball bearings;
- The inertia tensor of the mobile parts (including the corresponding uncertainties as described before);
- The leaf rigid-body and flexible modes, together with the modal damping;
- The contacts of the micro-switches;
- The misalignments in the RSU mounting and between the HS and MS rotation axes.

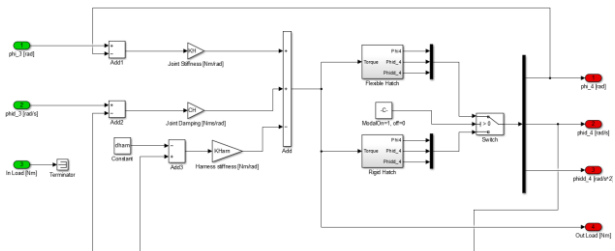

*Figure 9: RSU Leaf block diagram*

The analyses performed with this model permitted to determine all the budgets needed for the balancing, alignment, control precision, and also the influence of the RSU motion profiles.

These analyses have also shown that, among all contributors to the high frequency content of the exported loads, the hatch flexible modes are playing a dominant role. This is shown in Figure 11 (see the four green peaks).

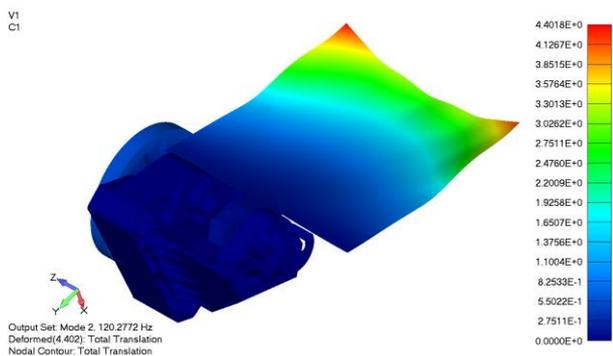

*Figure 10: The first RSU flexible mode at 120Hz.*

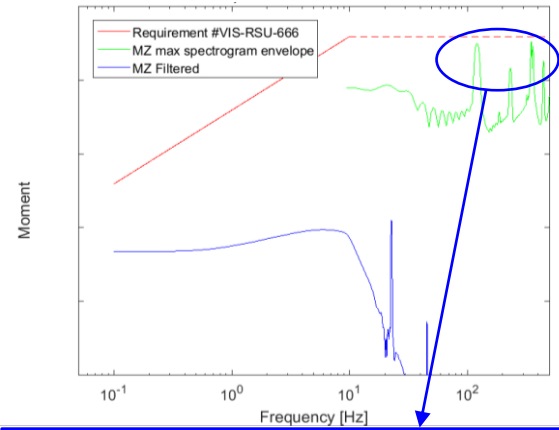

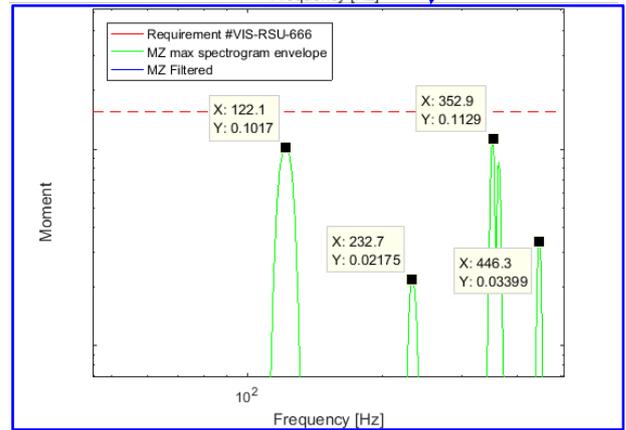

*Figure 11: Exported micro-vibrations induced by the RSU along the Z direction.*

**DESIGN EVOLUTIONS**

The RSU design has significantly evolved from the early baseline achieved at the end of phase B (see Figure 12) to the current phase C, mainly due to the increased complexity of the available micro-vibration model.

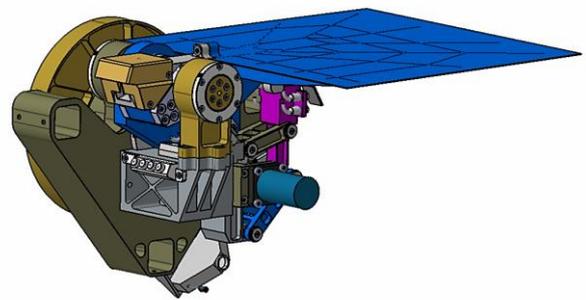

*Figure 12: The RSU design at the end of Phase B.*

The most relevant changes between the phase B and C design are:
- an improved configuration of the micro-switches, allowing us to increase the motorization margin;
- the suppression of the mobile Fail-Safe

Mechanism (FSM) on the HS, which was shown to degrade the unit reliability;
- the suppression of the Hold-Down and Release Mechanism (HDRM), as it was shown that the properly balanced MoIs of the two shafts and the stepper motor detent torque were sufficient to keep in place all the mobile parts even when subjected to the largest expected external loads;
- the introduction of all adjustment capabilities in order to keep the manufacturing feasibility at an acceptable level without degrading the micro-vibration performances;
- the introduction of an anti-backlash system in the gears to minimize the micro-vibrations experience during the first dedicated test of the bread-board model (see the next section).

The improved RSU design achieved at the manufacturing review of the flight model is shown in Figure 4.

**MODEL PHILOSOPHY**

The entire RSU development program foresees the production of the following different models:
- An electrical model (EM) to validate all electrical functions and interfaces of the units, verifying its ESD and EMC properties;
- A structural and thermal model (STM) to validate the test feasibility and to allow vibration and thermal tests at Spacecraft level;
- A Drive Train Model (DTM) to perform preliminarily micro-vibration tests and develop in house the required RSU control equipment (EGSE);
- A bread-board model (BM) to carry out the first full micro-vibration and lifetime tests, validating the RSU tribological elements, the driving motor and the micro-switches;
- A qualification model (QM);
- A flight model (FM);
- A flight spare (FS).

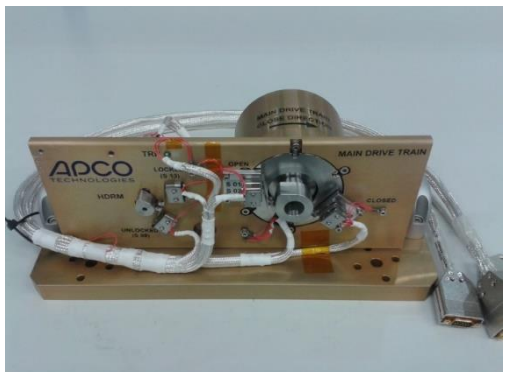
*Figure 13: The RSU Electrical Model.*

The RSU EM is representative of all the electronic components of the RSU. The harness is representative of the FM design in terms of lengths, wire gauges, and shielding. The mechanical parts have been procured to comply with the established flights standard.

The RSU STM is a simplified model designed to be representative of the FM predicted mass and moment of inertia (MoI) and being used for the structural and thermal tests. All the electrical components are replaced by mass dummies, while the rest (materials, coatings, paints, etc) is representative of the flight model.

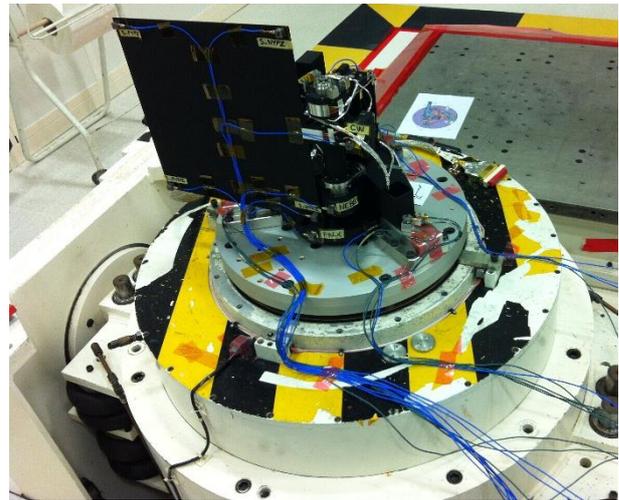
*Figure 14: the RSU Structural and Thermal Model.*

The RSU BM is used to perform the first measurements of the unit micro-vibration and exported torque performances (evaluated so far only through numerical modelling), together with a first lifetime test. In order to serve these objectives, the RSU BM was designed to include all contributors to the micro-vibrations and exported loads identified by analysis. It also requires that the HS and the MS are fully representative of those of the FM in terms of materials, geometry, lubrication, assembly, adjustment, and balancing procedure. Some geometrical simplification of the support structure is applied for programmatic reasons.

The RSU DTM is a simplified version of the BM that can be operated in ambient conditions, making use of wet lubricants and avoiding any paints.

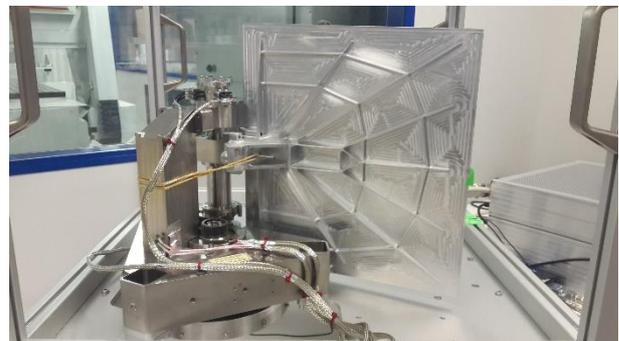
*Figure 15: The RSU Drive Train Model.*

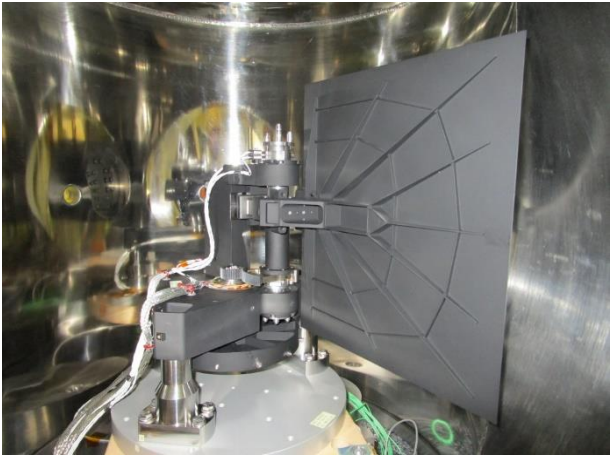

*Figure 16: The RSU Breadboard Model.*

## ASSEMBLY AND INTEGRATION

In addition to the usual installations and tools to perform assembly and integration of space mechanisms, 3 specific facilities have been used for the RSU.

The first one, mentioned earlier, is the precise Schenck's MoI measurement machine shown in Figure 17 and specifically procured for the RSU project.

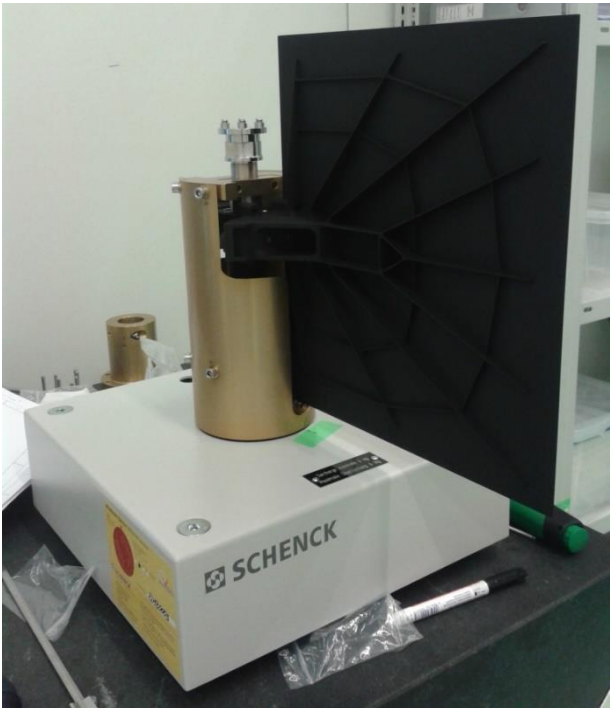

*Figure 17: The Schenck machine, here used to perform the precise measurement of the RSU Hatch Shaft MoI.*

The second one is a machine available at the Schenck premises that has been used to perform the static and dynamic balancing of the RSU shafts. The major problem that we encountered during these measurements was related to the impact of the air flow produced by the moving RSU leaf onto the machine frame. A "bell" has been used to limit this effect, as shown in Figure 18.

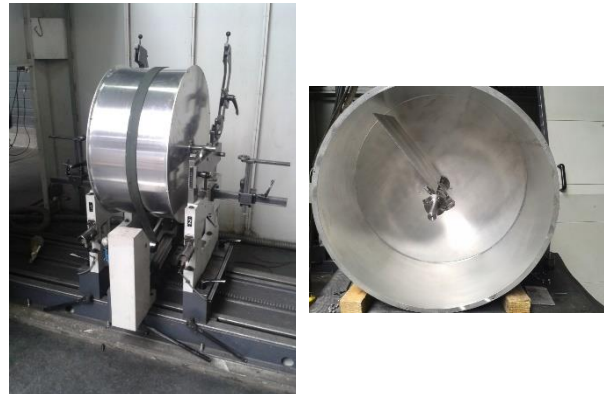

*Figure 18: Machines at Schenck premises used to the perform the balancing of the RSU Hatch Shaft.*

The last facility that has been critical for the APCO Technologies' activities was a vacuum chamber installed in house. The availability of this facility ensures that all actuations and run-in activities of the RSU can take place in a proper environment, preserving at best the dry lubricant applied on all unit tribological elements.

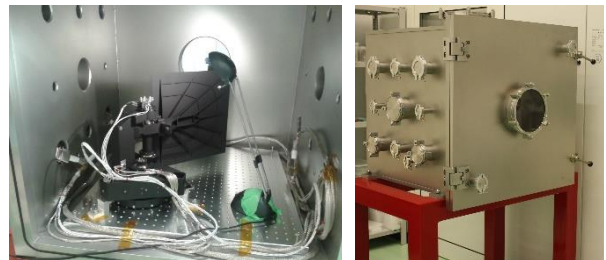

*Figure 19: The RSU BM inside the vacuum chamber installed at APCO's premises.*

## TEST

All the tests foreseen for the full qualification of the RSU have been performed at least once on the DTM, BM, STM, and EM, to de-risk the development of the most critical models (QM, FM, and FS).

The ESD and EMC tests have been successfully conducted on the EM, while the compliance of the unit to the foreseen vibration levels has been proven by using the STM and the BM. These tests have also been used to validate the removal of the HDRM. Reduced and complete micro-vibration tests have been carried out on the DTM and the BM, respectively.

The lessons learned from these test campaigns are summarized below.

**Vibration test**

The BM vibration test demonstrated that the HDRM was not needed. The wear at the mesh location was found to be virtually identical with or without the holding mechanism. An in-depth inspection with the Scanning Electron Microscope (SEM) was performed to check that the MoS2 layer was still present after the vibration test.

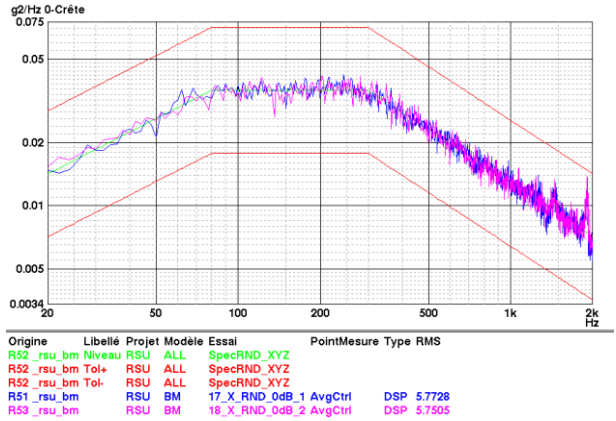

*Figure 20: Random vibration level passed successfully.*

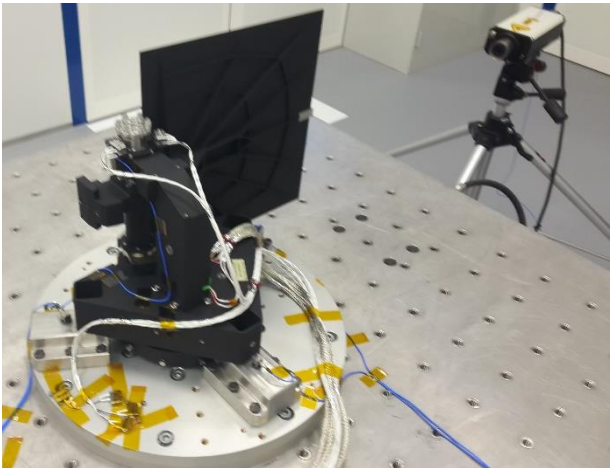

*Figure 21: RSU vibration test performed using a laser vibrometer to avoid accelerometers mounted on the moving parts.*

**Thermal vacuum test**

The target temperature of 110K is at the limit of the performances of vacuum chambers making use of gas nitrogen GN2 thermal generators. The test has been repeated two times to achieve the targeted temperature of 110 K with the GN2 available at Intespace. This facility allows us to cool down both the cold plate and the shrouds and heat losses were minimizedthrough the inclusion of an additional insulation layer. The experience gained through this test guarantees that a proper QM qualification campaign can be carried out at Intespace.

**Lifetime**

The BM successfully underwent a lifetime test during which 560'472 and 21 opening and closing sequences have been performed respectively in vacuum and in air. A new SEM inspection after the test revealed that the coating was in good conditions. In the zone worn by the vibration test, the coating was even more homogeneously distributed than before the lifetime test.

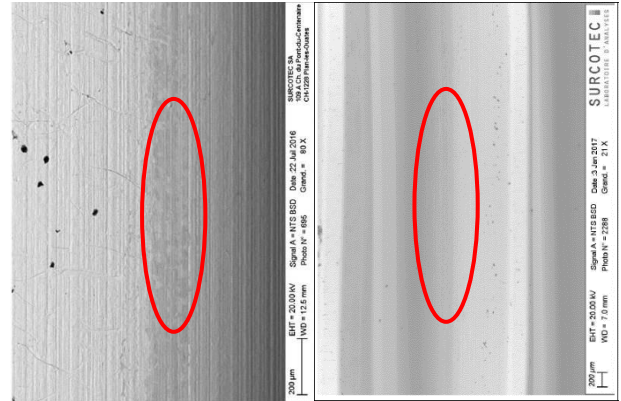

*Figure 22: The RSU gears SEM inspection before (left) and after (right) the lifetime test.*

**Microvibration**

The first complete microvibration test was performed at ESA/ESTEC on the Reaction wheel Characterisation Facility (RCF). The measured levels were largely above the expectations, as shown in Figure 23.

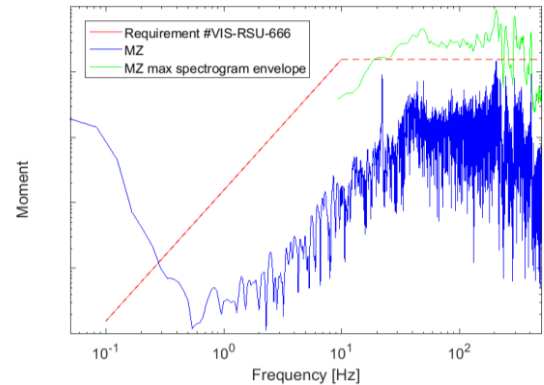

*Figure 23: The measured RSU Exported torque at RCF.*

This test highlighted an issue in the simulation with the implementation of the backlash function and related trade-off concluding that no anti-backlash system was necessary.

After the revision of the analysis, an anti-backlash has been designed and tested first on the DTM. The results obtained are promising as shown in Figure 24.

The design chosen will allow a late decision to implement or not anti-backlash system on the QM. A second lifetime test on the BM, including the anti-backlash system modification, will help in the decision.

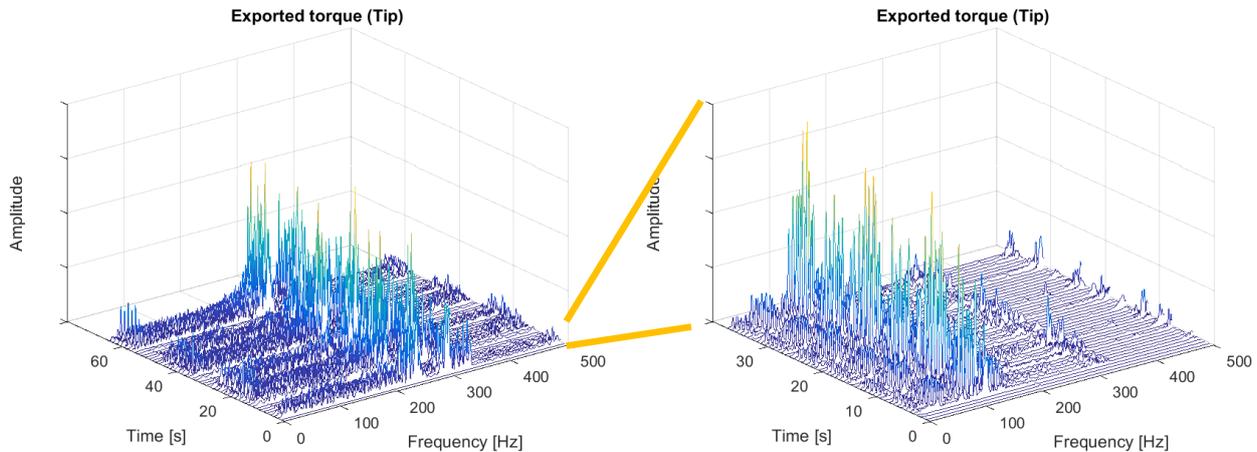

*Figure 24: The RSU DTM induced micro-vibrations measured without (left) and with (right) the anti-backlash system.*

## CONCLUSIONS

The RSU design has been continuously improved over the past 6 years to comply with the expected operations in a harsh environment and with all the challenging Euclid and VIS instrument requirements.

The RSU EM and STM programs have been completed and the two models successfully passed all foreseen tests.

The DTM has proved very helpful in designing the EGSE and performing intermediate tests.

The BM model has successfully confirmed the capability of the selected components to sustain the planned long-term operations at cryogenic temperatures in the nominal configuration.
In a second lifetime test to be carried out in late 2017, the BM will be used to confirm the compliance of the unit with the long-term operations when the anti-backlash system is included to minimize the micro-vibrations.

The design of the RSU is expected to reach full maturity at the end of 2017, when a fully de-risked QM test campaign will be initiated before the manufacturing of the FM and FS (starting in 2018).

## ACKNOWLEDGEMENT

The Euclid team at the University of Geneva and APCO Technologies acknowledges the support from the Swiss State Secretariat for Education, Research and Innovation (SERI) and ESA's PRODEX programme. We also thank all members of the Euclid consortium for the fruitful collaboration and the pleasant working environment. We are grateful to the ESA Prodex team for their extensive support and collaboration on the project.